# Directional massless Dirac fermions in a layered van der Waals material with one-dimensional long-range order

T. Y. Yang[1,#], Q. Wan[1,#], D. Y. Yan[2,3,#], Z. Zhu[4], Z. W. Wang[5], C. Peng[1], Y. B. Huang[6], R. Yu[5], J. Hu[7], Z. Q. Mao[8], Si Li[9], Shengyuan A. Yang[9], Hao Zheng[4,10], Jin -Feng Jia[4,10], Y. G. Shi[2,3,11] and N. Xu[1,*]

[1] *Institute for Advanced Studies, Wuhan University, Wuhan 430072, China*
[2] *Beijing National Laboratory for Condensed Matter Physics and Institute of Physics, Chinese Academy of Sciences, Beijing 100190, China*
[3] *School of Physical Sciences, University of Chinese Academy of Sciences, Beijing 100190, China*
[4] *Key Laboratory of Artificial Structures and Quantum Control (Ministry of Education), Shenyang National Laboratory for Materials Science, School of Physics and Astronomy, Shanghai Jiao Tong University, Shanghai 200240, China*
[5] *School of Physics and Technology, Wuhan University, Wuhan 430072, China*
[6] *Shanghai Synchrotron Radiation Facility, Shanghai Institute of Applied Physics, Chinese Academy of Sciences, Shanghai 201204, China*
[7] *Department of Physics, University of Arkansas, Fayetteville, AR 72701, USA.*
[8] *Department of Physics, Pennsylvania State University, University Park, PA 16802, USA.*
[9] *Research Laboratory for Quantum Materials, Singapore University of Technology and Design, Singapore 487372, Singapore*
[10] *Tsung-Dao Lee Institute, Shanghai Jiao Tong University, Shanghai, 200240, China.*
[11] *Songshan Lake Materials Laboratory, Dongguan, Guangdong 523808, China*

\# These authors contributed equally to this work.
\* E-mail: nxu@whu.edu.cn

**Exotic properties in single or few layers of van der Waals materials carry great promise for applications in nanoscaled electronics, optoelectronics and flexible devices. The established, distinct examples include extremely high mobility and superior thermal conductivity in graphene, a large direct band gap in monolayer MoS$_2$ and quantum spin Hall effect in WTe$_2$ monolayer, etc. All these exotic properties arise from the electron quantum confinement effect in the two-dimensional limit. Here we report a novel phenomenon due to one-dimensional (1D) confinement of carriers in a layered van der Waals material NbSi$_{0.45}$Te$_2$ revealed by angle-resolved photoemission spectroscopy, i.e. directional massless Dirac fermions. The 1D behavior of the carriers is directly related to a stripe-like structural modulation with the long-range translational symmetry only along the stripe direction, as perceived by scanning tunneling microscopy experiment. The four-fold degenerated node of 1D Dirac dispersion is essential and independent on band inversion, because of the protection by nonsymmorphic symmetry of the stripe structure. Our study not only provides a playground for investigating the striking properties of the essential directional massless Dirac fermions, but also introduces a unique monomer with 1D long-range order for engineering nano-electronic devices based on heterostructures of layered van der Waals materials.**

The layered van der Waals (vdW) materials have become one of central topics in condensed-matter physics and materials science since the successful exfoliation of graphene [1]. Due to quantum size effect of electrons in the two-dimensional (2D) limit, vdW materials with nanosized thickness display qualitative distinctions from the bulk compounds, including the 2D massless Dirac fermion in graphene with extremely high mobility [1], the large and direct band gap in single layered $MoS_2$ [2], Ising superconductivity in $MoS_2$ and $NbSe_2$ [3-4], quantum spin Hall effect in $WTe_2$ monolayers [5-9] and 2D ferromagnetism in single layered $CrI_3$ [10-11], etc. Furthermore, due to the 2D crystal structural geometry, atomically thin layers of vdW materials can be assembled into heterostructures with high designability and manipulability, presenting emergent behaviors, for example superconductivity in bi-layered graphene with magic angle [12-13]. The novel optical, electronic, and mechanical properties in layered vdW materials and heterostructures hold great potential applications in the next generation of electronics and optoelectronics in nanoscale.

Here, we present a direct experimental evidence that $NbSi_{0.45}Te_2$, a vdW material with layered structure, exhibits one-dimensional (1D) confinement of its metallic surface state and yields novel directional massless Dirac fermions, as probed by angle-resolved photoemission spectroscopy (ARPES) measurements. The 1D behavior originates from the remarkable stripe-like surface, that only has long-range translation symmetry along the stripe direction as directly visualized by scanning tunneling microscopy (STM) experiments. The nonsymmorphic symmetry of the stripe structure guarantee the four-fold degeneracy of the 1D Dirac node, make the directional massless Dirac fermions essential [14-15] and distinct from ones rely on band inversion [16]. Our study not only experimentally reveals the directional massless Dirac fermions which is the 1D analogue to the 2D/3D ones in graphene/Dirac semimetal, but also provides a unique layered vdW monomer with 1D long-range order, for constructing functional device based on 2D heterostructures.

$NbSi_xTe_2$ is composed of stacked Te-$(NbSi_x)$-Te sandwich layers, that can be considered as intercalation of Si into a transition metal dichalcogenide (TMD). Distinct from other cases of reaction of Fe (or Ni or Co) with $(Nb,Ta)Te_2$ [17-20], that leads to a totally new phase with structure unrelated to the parent compounds, $NbSi_xTe_2$ exhibits similar structure properties as the 2H-TMD (Fig. 1a-I). In NbSi$_{1/3}$Te$_2$, as seen from Fig. 1a-II, the deformation induced by the intercalated Si atoms make the original hexagonal structure of NbTe$_2$ layer transform into an square plane [21]. Si atoms adopt a Te square environment in the center of the empty biprisms (Fig. 1a-III). NbSi$_{1/3}$Te$_2$ is effectively built from zigzag chains with the (Nb-Nb)-(Si) cationic sequences (chains **a** and **b** in Fig. 1b-I) and the NbTe$_2$ chains (chain **c** in Fig. 1b-I), with **abc** chain succession (Fig. 1b-II). The interlayer Te-Te distance is in the range of typical van der Waals type and atomically thin NbSi$_{1/3}$Te$_2$ flakes were successfully fabricated by micro-exfoliation [22], that demonstrates the capability as a monomer for nanoelectronics based on van der Waals heterostructure. Topological semimetal phases are proposed in both bulk and single-layered NbSi$_{1/3}$Te$_2$ by first principle calculations study [23]. As the Si intercalation level increasing, more Nb atoms are paired and form the **a**(**b**) type chains. Because the Nb-Si bonding is necessary to stabilize the **c** type chains, two **c** chains are forbidden to be side by side [24]. Therefore, the NbSi$_x$Te$_2$ phase is composed by (**ab**)×n-**c** type structure (Fig. 1b-III) with the relationship of $x = n/(2n+1)$. A longer periodicity along the y direction is expected for larger x value, for example **ababc** for x = 2/5 and **abababc** for x = 3/7. Correspondingly, the Brillouin zones (BZ) shrinks along the k$_y$ direction, as illustrated by the (001) surface BZs for x = 1/3, 2/5 and 3/7 (n = 1, 2 and 3) shown in Fig. 1c. The highest Si intercalation level is x = 1/2 in NbSi$_{1/2}$Te$_2$ (Fig. 1a-IV), that corresponds to the (**ab**) structure without any **c** chain (Fig. 1b-IV).

High quality NbSi$_{0.45}$Te$_2$ single crystals in a few millimeters size were synthesized by flux method, which are suitable for electronic structure investigation by ARPES. The stoichiometry is determined by energy dispersive x-ray spectroscopy. In Fig. 1d, we present the photoemission intensity in a long range of binding energy (E$_B$), in which Si *2p* core level peaks, together with ones of the Nb *4p* and Te *4d* electrons, are clearly observed, confirming the successful intercalation of Si atoms in our single crystal samples. The core level data is compared with that obtained from NbSi$_{1/3}$Te$_2$, with the results shown in the insets of Fig. 1d. The Si-*2p* peaks for x = 0.45 sample (red line in the inset-I of Fig. 1d) are more intensive than ones for x = 1/3 sample (blue line in the inset-I), indicating a higher Si intercalation level that is fully consistent with the energy dispersive x-ray spectroscopy results. For the Te-4d electrons, the shoulder-like features at lower E$_B$ are strongly suppressed and the main peaks at higher E$_B$ are enhanced in x = 0.45 sample, with the total area of Te-4d peaks unchanged. The shoulder-like features correspond to Te atom chains in the **c** chains that do not contain the intercalated Si atoms (indicated by the red dashed squares in Fig. 1a-b). Our core level results (Fig. 1d) is fully consistent with the scenario shown in Fig, 1b, that the increasing of Si intercalation level x reduces the ratio between the number of chain c to that of chain a (or b).

To verify the bulk/surface origin of the ARPES signals, we performed photon energy ($h\nu$) dependent measurement along the Γ-X direction, with the constant energy plot at $E_F$ shown in Fig. 2a. The most pronounced feature of the raw data shown in Fig. 2a is the sudden intensity suppression at $h\nu$ around 32 eV, that corresponds to the anti-resonance between the Nb 4p state to valence state near $E_F$ (Fig. 1d). Away from the anti-resonance energy range, the photoemission intensity of the valence states recovers and further becomes weaker at higher $h\nu$ due to the smaller photoemission cross-section of Nb 4d electrons. Although the intensity is modulated with *hv* in Fig. 2a, the momentum positions of the experimentally observed features show no variations with *hv*, suggesting a 2D Fermi surface (FS) configuration. In order to confirm the 2D nature of the observed valence bands near $E_F$, we plot the *hv* dependence spectra taken at the $\bar{\Gamma}$ and $\bar{X}$ points of the surface BZ in Fig. 2b and c (indicated by cut 1 and cut2 in Fig. 2a), with the corresponding energy distribution curves (EDCs) shown in Fig. 2d and e, respectively. The spectra taken with different *hv* shown in Fig. 2b-e are normalized to total intensity, to eliminate intensity variation caused by photoemission cross-section and only focus on the band dispersions. Our results clearly indicate that there are no observable dispersions along the $k_z$ direction for the experimentally determined electronic states at the $\bar{\Gamma}$ point, with one peak at $E_B$ ~ 30 meV and another broad peak at $E_B$ ~ 300 meV (Fig. 2b and d). Similarly, two non-dispersing bands are observed around 100 meV and 600 meV below $E_F$ at the $\bar{X}$ point (Fig. 2c and e).

On the other hand, similar to the parent compound NbSe$_2$ [25], TaSi$_{1/3}$Te$_2$ [26] and other vdW TMDs [27], electrons are expected to be coherently hopping between NbSi$_x$Te$_2$ layers, and forms 3D FSs and band structure dispersing along the $k_z$ direction [23]. Therefore, the electronic state without $k_z$ dispersing, as observed in our ARPES experiments here, is corresponding to the surface state. The surface state observed here originates from an uncommon stripe-like surface, which will be discussed more in details below.

We plot the surface band structure along the $\overline{\Gamma}$-$\overline{X}$ direction in Fig. 2f, probed by ARPES using photon energy hv = 48 eV, with the corresponding curvature and EDC plots shown in Fig. 2g and h, respectively. We observed a hole-like band, named α, with a relatively flat band top slightly below $E_F$ near the $\overline{\Gamma}$ point (corresponding to the $E_B$ ~ 30 meV peak shown in Fig. 2b and d). Another band, named β, shows an almost linear dispersion and crosses $E_F$ near the surface boundary $\overline{X}$ point. Here we note that a band β' (indicated by dashed line in Fig. f) is expected to appear and be symmetric to the β band along the BZ boundary $\overline{X}$, due to the translation symmetry along the x direction. The photoemission spectra weight of the β' band is strongly suppressed along the $\overline{\Gamma}$-$\overline{X}$ direction. This phenomenon, that photoemission from a certain band is observed in only every other BZ, was previously reported in nonsymmorphic systems [27-30], because of a selection rule directly related to the non-primitive translation operations [28]. Here the absence of the β' band along the $\overline{\Gamma}$-$\overline{X}$ direction is induced by the nonsymmorphic symmetry of the stripe region in NbSi$_{0.45}$Te$_2$. In fact, the same nonsymmorphic symmetry $\widetilde{M_y}$ protects the essential four-fold degeneracy point of the 1D Dirac dispersion observed near the stripe region in NbSi$_{0.45}$Te$_2$, together with the mirror and time reversal symmetry, which will be discussed later.

We map the surface states in the $k_x$-$k_y$ plane, with the FS intensity and the corresponding second derivative plotted in Fig. 2i and j, respectively. The surface BZs for the (**ab**)×n-**c** structures with several n values are appended. Surprisingly, the observed FSs, especially the β (and the β') FS near the BZ boundary, show a clear 1D behavior along the $k_y$ direction within the whole region of surface BZs for various of n values, that is very similar to the experimentally determined FS in $k_x$-$k_z$ plane (Fig. 2a). The out-of-plane and in-plane FS results (Fig. 2a and k) together demonstrate a 1D FS topology of the β band, that shows no dispersion along both the $k_y$ and $k_z$ directions.

Not only the FS topology, in fact the whole β band shows a 1D Dirac dispersion with an unusual Dirac point in the momentum space, as demonstrated by the results shown in Fig. 3. At the $k_y$ value corresponds to the BZ boundary of NbSi$_{1/3}$Te$_2$ with n = 1 (the $\overline{Y}_1$-$\overline{U}_1$ direction in Fig. 1c), both the β band and the folding band β' can be clearly observed in the ARPES intensity plot (Fig. 3a) and corresponding curvature plot (Fig. 3b), because the selection rule related to the nonsymmorphic symmetry is not strictly applied off the $\overline{\Gamma}$-$\overline{X}$ direction. The β and β' bands cross each other along the $k_x$ direction and form a Dirac dispersion, with the Dirac point around 0.1 eV below $E_F$ with four-fold degeneracy. Such a Dirac dispersion displays a 1D behavior, as indicated by the ARPES intensity in the surface BZ at a serial of the constant $E_B$ shown in Fig. 3c and the corresponding curvature in a zoomed region (indicated by red square in Fig. 3c) plotted in Fig. 3d. When the energy shifts from $E_F$ to higher $E_B$ (cut$_1$-cut$_3$ in Fig. 3b), two straight lines (cut 1), corresponding to the 1D β and β' FSs, firstly merge into a single line at $E_B \sim 0.1$ eV (cut 2) and then separate into two lines again (cut 3). The evolution of the β and β' bands is fully consistent with a 1D Dirac dispersion scenario. Figure 3e and f present the ARPES intensity and curvature along the $\overline{X}$-$\overline{U}$ high symmetry line, respectively, in that a single flat band is observed corresponding to the 1D Dirac point. Considering the results shown in Fig. 3e and Fig. 2c together, we provide unambiguous experimental evidence of 1D Dirac dispersion hosted on the surface of NbSi$_{0.45}$Te$_2$, that the Dirac point shows no dispersing along both the $k_y$ and $k_z$ directions. We note the α band shows a dispersive feature along the $k_y$ direction in Fig. 3d. As discussed in Supplementary Information, the observed periodicity of 1.5 Å$^{-1}$ relates well to the inter-block spacing within the (**ab**)×n ribbons of $b_0 = 4.1$ Å. It is attributed to the intra-ribbon dispersion, similar as the intra-molecule dispersion observed in sexiphenyl previously [31].

The unique 1D Dirac dispersion experimentally observed here is derived from an abnormal stripe-like modulation, which is directly visualized by STM experiments. As seen from Fig. 3g, stripes with alignment along the x direction are directly visualized on the NbSi$_{0.45}$Te$_2$ surface. This stripe-like pattern is a direct consequence of the **c** type chain component in the structure of NbSi$_x$Te$_2$ as shown in Fig. 1. Because of the weak vdW interaction between the layers in NbSi$_x$Te$_2$, the surface probed by the STM experiment is terminated by Te atoms (Fig. 1a-III). The Te atoms in **c** type chains shows a different chemical environment from others due to the absent of Si intercalations in chain **c** (Fig. 1a-III and b). It can induce valence and height variations of Te atoms near the **c** type chain region, and results in the stripe-like contrast in STM images (Fig. 3g). Interestingly, in a typical scale of in 100 nm × 100 nm area, the distance between the stripes does not favor a fixed value, but varies from 3.43 nm to 5.4 nm (Fig. 3g-h). On the other hand, the lattice periodicity along the stripe direction keeps the value of 6.35 Å (Fig. 3i-j), that corresponds to the bulk lattice parameter *a* along x direction. Therefore, our STM results on the NbSi$_{0.45}$Te$_2$ surface point to an unconventional stripe-like structural modulation, in which a certain inter-stripe distance is absent and the long-range translational symmetry is only reserved along the stripe direction (the x axis).

Such an unusual stripe-like structural modulation with 1D long-range order is a unique property of NbSi$_{0.45}$Te$_2$ system. As discussed above, the distance between the nearby bright stripes is the total width of the unit of (**ab**)×n-**c**, that is directly related to the Si stoichiometry x with the relationship x = n/(2n+1). The inter-stripe distances of 3.43 nm, 4.30 nm and 5.15 nm in Fig. 3g-h correspond to the (**ab**)×n-**c** units with n = 4, 5 and 6, with the Si intercalation level x = 4/9, 5/11 and 6/13, respectively. The stoichiometry difference of Si between two successive n values is very tiny, that is 1% between n = 4 and 5 stripes and 7‰ between n = 5 and 6 stripes. In fact, the stoichiometry difference could be even smaller when both n-1 and n+1 stripes emerge together. In the case shown in Fig. 3g, the difference is 3‰ between a pair of stripes with n = 4,6 and two stripes with n = 5. Such a tiny local stoichiometry fluctuations are hardly, if not impossible, avoidable in the synthesizing process of the complex compounds with more than 100 atoms in (**ab**)×5-**c** unit ribbon. The fluctuations induce the unusual stripe-like structural modulation with 1D long-range order in NbSi$_{0.45}$Te$_2$, which in fact exhibits a novel type of atom arrangement [32] distinct from crystalline, quasicrystalline and amorphous.

The observation of the unique stripe structure with 1D long-range order provides a natural explanation of the 1D Dirac dispersion detected in ARPES experiment. As indicated by the first principles calculations in Supplementary Information, the Dirac dispersion formed by the β and β' bands is spatially localized around the **c** type chain regions which correspond to the stripes in STM results. As the spacing blocks between stripes become more than two sets of (**ab**) chains, the inter-stripe hopping term can be ignored and the Dirac dispersion near E$_F$ shows a quasi-1D behavior. In NbSi$_{0.45}$Te$_2$, the stripe intervals vary from four to six sets of (**ab**) chains. Therefore, the 1D Dirac dispersion stems on the surface of vdW material with 1D long-range order.

The four-fold degeneracy of the 1D Dirac dispersion observed in NbSi$_{0.45}$Te$_2$ is essential and protected by the nonsymmorphic symmetry of the stripe region (Fig. 4a), together with the mirror and time reversal symmetries. As shown in Fig. 4b-c, the crystal structure of the stripe region has a glide mirror symmetry $\widetilde{M_y}$ perpendicular to the y direction $\widetilde{M_y}$ (x,y,z)→(x+1/2,-y,z). The $\widetilde{M_y}$ corresponds to a non-primitive translation operation along the x direction, with eigenvalues of $g_{\widetilde{M_y}} = \pm \lambda e^{-ika/2}$. The additional phase term in $g_{\widetilde{M_y}}$ makes eigenstates in a 4π periodicity, that corresponds to two BZs. Figure 4e display a generous case of band structure along the invariant line of $\widetilde{M_y}$, in that the dashed line is the folded band from 2$^{nd}$ BZ (k - 2π/a), with an additional phase term ($e^{-i\pi}$ = -1) in $g_{\widetilde{M_y}}$ compare to the main band. For NbSi$_{0.45}$Te$_2$, all the k points sit on the invariant line of $\widetilde{M_y}$ because of the 1D nature. Spin orbit coupling has to be included, therefore $\lambda$ takes value of $i$ and $g_{\widetilde{M_y}} = \pm i e^{-ika/2}$. Furthermore, time reversal symmetry $T$ is reserved in NbSi$_{0.45}$Te$_2$, that induces Kramer's pairs of bands (|u⟩ and $T$|u⟩ indicated by blue and red colors in Fig. 4g, respectively) degenerated at Kramer's points (k = 0 and π/a), for both the main and folded bands. Because of the commute relation between $\widetilde{M_y}$ and $T$,

$$\widetilde{M_y}T = T\widetilde{M_y},$$

values of $g_{\widetilde{M_y}}$ for Kramer's pairs are complex conjugate to each other. Therefore, at Kramer's points, the bands forming Kramer's pairs have different values of $\widetilde{M_y}$ = $\pm i$ at k = 0 and same values of $\widetilde{M_y}$ = 1 (or -1) at k = π, as indicated in Fig. 4g. Finally, another mirror plane perpendicular to the z direction M$_Z$(x,y,z)→(x,y,-z) is reserved in the structure of the stripe region (Fig. 4d). Similar to the case of $\widetilde{M_y}$, all the k points in NbSi$_{0.45}$Te$_2$ are invariant under the symmetry operation of M$_Z$. Because $\widetilde{M_y}$ is anti-commute to M$_Z$ in the spin orbit coupled NbSi$_{0.45}$Te$_2$,

$$\widetilde{M_y}M_Z = -M_Z\widetilde{M_y},$$

the degenerated eigenstates |u⟩ and M$_Z$|u⟩ have opposite eigenvalues,

$$\widetilde{M_y}|u\rangle = g|u\rangle,$$
$$\widetilde{M_y}M_Z|u\rangle = -M_Z\widetilde{M_y}|u\rangle = -gM_Z|u\rangle,$$

In another word, each eigenstate |u⟩ with $g_{\widetilde{M_y}}$ = g is degenerated with its partner M$_Z$|u⟩ with $g_{\widetilde{M_y}}$ = -g, as shown in Fig. 4 g. Consequently, an Dirac point with four-fold degeneracy appears at k = π/a (Fig. 4g), that is guaranteed by the combination of $\widetilde{M_y}$, M$_Z$ and $T$

The 1D Dirac dispersion with four-fold degeneracy observed in NbSi$_{0.45}$Te$_2$ is essential [14-15] and distinct from those created by band inversion [16,33-35]. The essential Dirac node is formed by a pair of bands with additional phase term ($e^{-i\pi}$ = -1) in eigenvalues of the nonsymmorphic symmetry operation ($\widetilde{M_y}$ in NbSi$_{0.45}$Te$_2$), therefore band inversion is not indispensable to form the Dirac nodes. Due to the independence of band inversion, the only way to destroy the essential Dirac node is to lower the symmetries, then the system can transit to either topological or trivial insulator phase [14]. For the other distinct type of Dirac nodes observed in $Na_3Bi/Cd_3As_2$ [16, 33-35], a band inversion is necessary and a rotational symmetry guarantees the node with four-fold degeneracy. In this case, breaking the symmetry will open a gap at the Dirac nodes and induce a topological insulator phase. The trivial insulator is not adjacent to this kind of Dirac nodes related to band inversion. The nonsymmorphic symmetry protected Dirac nodal-line [36,37] and Weyl node [38,39] have been observed in previous studies, which correspond to the 2D and 3D Dirac dispersions, respectively, and are distinct from the 1D case as seen in $NbSi_{0.45}Te_2$ (see Fig. S3 in Supplementary Information).

Our experiment thus points to a 1D confinement of FS configuration on the surface of $NbSi_{0.45}Te_2$, induced by intercalation Si atoms into the Nb-Si plane of TMD $NbTe_2$. They provide a direct demonstration of the realization of 1D massless Dirac fermion, which forms the complete group of massless Dirac fermions classified with different dimensionalities together with the 2D one in graphene [1] and topological nodal-line semimetal [36,40-42], and the 3D ones in $Na_3Bi/Cd_3As_2$ [16,33-35]. Our results indicate that the directional massless Dirac fermions are directly related to the unique stripes in $NbSi_{0.45}Te_2$ with 1D long-range order, and are protected by the nonsymmorphic symmetry. Our experimental demonstration of 1D massless Dirac fermions not only opens a wealth of new opportunities for studying and controlling of 1D massless Dirac fermion, but also provides a unique piece of component with 1D long-range order to the tool box of nanoelectronics based on vdW heterostructures.

Phys. Rev. Lett. **117**, 016602 (2016).

## Acknowledgements

This work was supported by the National Key R&D Program of China (Grant No. 2018FYA0305800, 2016YFA0300403, 2017YFA0302901), the Ministry of Science and Technology of China (2018YFA0307000), the National Natural Science Foundation of China (Grant No. 11874047, 11674226, 11790313, 11774399), the Fundamental Research Funds for the Central Universities (Grant No. 2042018kf-0030), Beijing Natural Science Foundation (Z180008) and the K. C. Wong Education Foundation (GJTD-2018-01). Z.Q.M. acknowledges the support by the US Department of Energy under grants DE-SC0019068. N.X. acknowledges support by Wuhan University startup funding.

## Author contributions

N.X. conceived the experiments. T.Y.Y, Q.W., C.P. and N.X. performed ARPES measurements with the assistance of Y.B.H. D.Y.Y and Y.G.S. synthesized the NbSi$_{0.45}$Te$_2$ single crystals. Z.Z, H.Z. and J.F.J. performed STM measurements. Z.W.W, R.Y., S.L., S.A.Y. performed *ab initio* calculations. H.J. and Z.Q.M synthesized the NbSi$_{1/3}$Te$_2$ single crystals as reference samples. T.Y.Y, Q.W., H.Z. and N.X. analyzed the experimental data. N.X. wrote the manuscript. All authors discussed the results and commented on the manuscript.

## Method

Single crystals of NbSi$_{0.45}$Te$_2$ were grown by using Te as flux. Starting materials Nb (Powder, 99.99%, Alfa Aesar), Si (Lump, 99.9999%, Alfa Aesar) and Te (Lump, 99.999%, Alfa Aesar) were mixed in an Ar-filled glove box at a molar ratio of Nb:Si:Te = 3:1:30. The mixture was placed in an alumina crucible, which was then sealed in an evacuated quartz tube. The tube was heated to 1100 °C over 10 h and dwelt for 20 h. Then, the tube was slowly cooled down to 800 °C at a rate of 2 °C/h followed by separating the crystals from the Te flux by centrifuging. Shiny crystals with the size of 2×2 mm$^2$ and thick ness of tens of micron were obtained on the bottom of the crucible. Clean surfaces for ARPES measurements were obtained by cleaving NbSi$_{0.45}$Te$_2$ samples *in situ* in a vacuum better than 5×10$^{-11}$ Torr. Surface-sensitive VUV-ARPES measurements were performed with a Scienta Omicron DA30L analyzer, with an overall energy resolution of the order of 20 meV at $T$ = 30 K. The beam spot size in ARPES measurements is tens of microns. The STM measurements were conducted in a Unisoku STM (1600) at $T$ = 4.8 K. The samples were cleaved *in situ* at room temperature under ultrahigh vacuum and then loaded into the STM head immediately. Electrochemically etched tungsten tips were used after in situ electron-beam cleaning and treated on a clean Ag (111) substrate.

## Data availability

The data that support the plots within this paper and other findings of this study are available from the corresponding author upon reasonable request.

# Figures

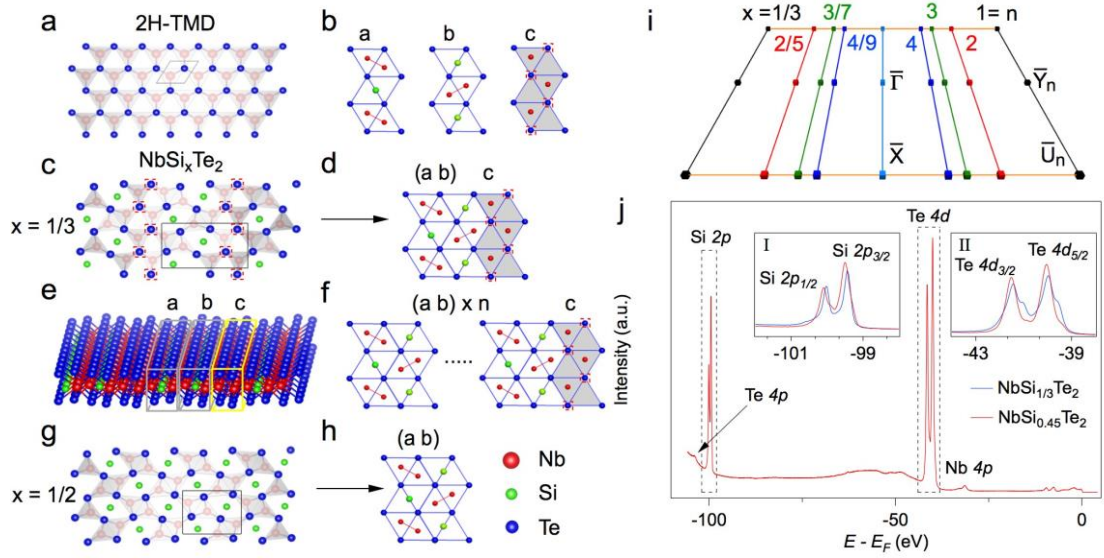

**Figure 1. Crystal structure and core-level spectra of NbSi$_x$Te$_2$. a,** Crystal structure of 2H-phase of transition metal dichalcogenid (2H-TMD) single-layer. The light shadows indicate the coordination polyhedral. **b**, The schematic building units of NbSi$_x$Te$_2$. The red, green and blue balls represent Nb, Si and Te atoms, respectively. The **a/b** units are zigzag chains with the (Nb-Nb)-(Si) cationic sequences, and the **c** unit is NbTe$_2$ chains. The blue balls marked by red squares represent the Te atoms in **c** type chains with a different chemical environment from others. **c-e,** Crystal structure, schematic and 3D presentation of NbSi$_{1/3}$Te$_2$ single-layer, which is composed by **abc** chains succession. **f,** The schematic of NbSi$_x$Te$_2$ single-layer, which is in a (**ab**)×n-**c** type structure with a relationship of x = n/(2n+1). n is the number of (**ab**) blocks between adjacent **c** chains. **g,h,** Crystal structure and schematic of NbSi$_{1/2}$Te$_2$ single-layer, which is composed by **ab** chains succession. **i,** Surface Brillouin zone of NbSi$_x$Te$_2$ with different x values. Corresponding n values for NbSi$_x$Te$_2$ with (**ab**)×n-**c** structure are also indicated. The light blue line in the center represents the $\bar{\Gamma}$-$\bar{X}$ high symmetry line. The black, red, green and blue lines correspond to $\bar{Y}_n$-$\bar{U}_n$ high symmetry lines for n = 1, 2, 3 and 4, respectively. The orange lines represent the $\bar{X}$-$\bar{U}_n$ high symmetry line. The high symmetry points $\bar{\Gamma}$, $\bar{X}$, $\bar{Y}_n$ and $\bar{U}_n$ are indicated by the dots with same colors as the high symmetry lines they located on. **j,** Photoemission spectra in a wide energy range from Fermi energy (E - E$_F$) for core levels of NbSi$_{0.45}$Te$_2$ (red lines) and NbSi$_{1/3}$Te$_2$ (blue lines). Insets I and II are close-ups of the Si *2p* and Te *4d* levels, respectively.

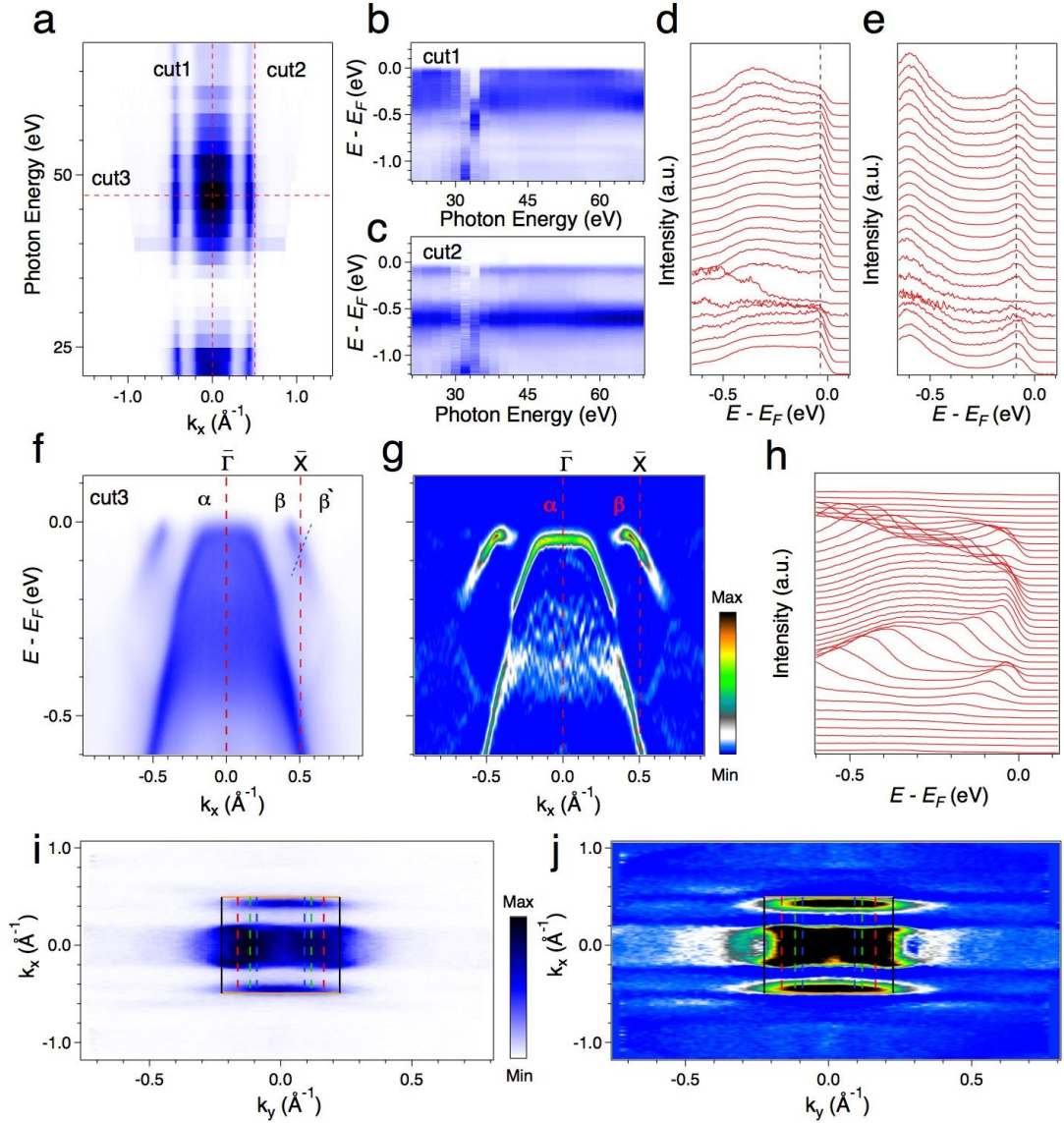

**Figure 2. Surface electronic structure of NbSi$_{0.45}$Te$_2$. a,** Plot of raw photoemission data at Fermi energy (E$_F$) along the $\bar{\Gamma}$-$\bar{X}$ direction. **b,c,** Normalized photoemission intensity plots at the $\bar{\Gamma}$ and $\bar{X}$ points, as indicated by cut1 and cut2 in **a**, respectively. **d,e,** Corresponding energy distribution curve (EDC) plots. **f-h,** Photoemission intensity, curvature and EDC plots along the $\bar{\Gamma}$-$\bar{X}$ direction, indicated by cut3 in **a**. Bands near E$_F$ are labelled as α and β. The dashed line in **f** indicates the β' band, which is guaranteed by the translation symmetry along the x direction to be symmetric to the β band along the BZ boundary $\bar{X}$. The spectra weight of the β' band is strongly suppressed along the $\bar{\Gamma}$-$\bar{X}$ high symmetry line due to photoemission selection rule. **i,j,** The photoemission intensity plot and the corresponding curvature plot at E$_F$ in the k$_x$-k$_y$ plane, respectively. The boxes indicate the surface BZs for the (**ab**)×n-**c** structures with several n values, which is the same as **Fig. 1i**.

Page 18

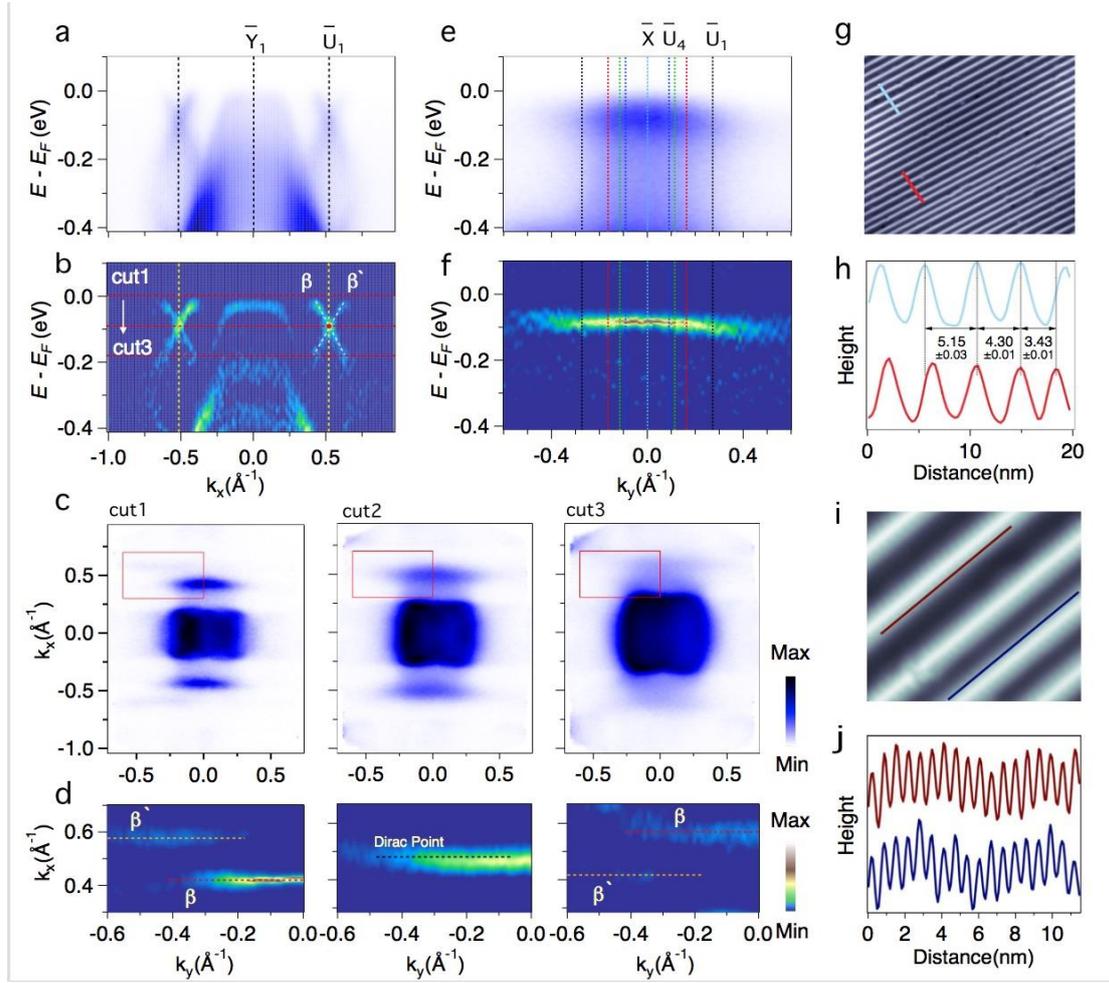

**Figure 3. 1D Dirac dispersion on stripe-like surface of NbSi$_{0.45}$Te$_2$. a,b,** Photoemission intensity plot and the corresponding curvature plot along the $\overline{Y}_1$-$\overline{U}_1$ direction, respectively. **c,d,** The photoemission intensity plot and the corresponding curvature plot in a zoomed in area at different binding energy (E$_B$), respectively. The E$_B$ of each panel is indicated by cut1-cut3 in **b**. The dashed lines are the guide for the eye of the β and β' bands. **e,f** Same as **a,b**, but along the $\overline{X}$-$\overline{U}_n$ direction. The dashed lines indicate high symmetry points $\overline{X}$ and $\overline{U}_n$, with same colors as **Fig. 1i**. **g,h**, A large scale scanning tunneling microscopy (STM) image (100×100 nm$^2$, 400mV, 1nA) with two line profilers clearly demonstrating the various values of inter-stripe distances (5.15±0.03 nm, 4.30±0.01 nm and 3.43±0.01 nm). **i,j,** Atomic scale STM image (14×14nm$^2$, 300mV, 0.5nA) with two line profiler showing the periodical lattice along stripes with different inter-stripe distances.

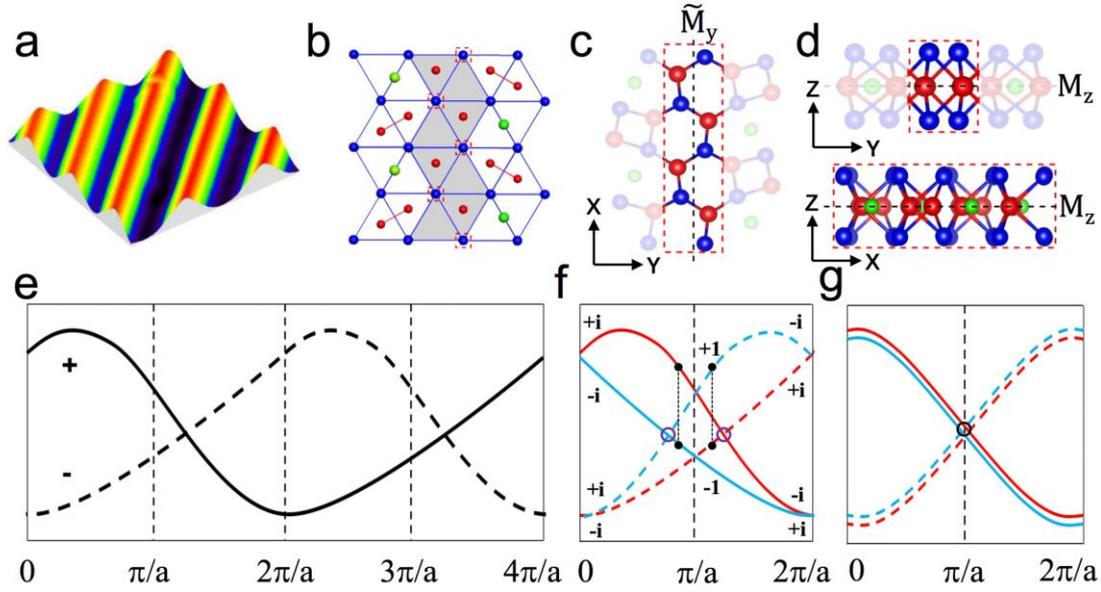

**Figure 4: The nonsymmorphic symmetry protected 1D Dirac dispersion with four-fold degeneracy. a**, The three-dimensional presentation of STM results on the stripe surface of NbSi$_{0.45}$Te$_2$. **b,c**, The illustration and crystal structure near the **c** type chain in NbSi$_{0.45}$Te$_2$ projected along the z direction, respectively. The glide mirror ($\widetilde{M}_y$) is indicated by the black dashed line. **d**, The crystal structure of the **c** type chain region projected along the x and y directions. The mirror plane ($M_z$) is marked by the black dashed lines. The **c** type chain is highlighted by shadow in **b** and red dashed square in **c** and **d**. **e**, Band structure along the invariant line of glide mirror symmetry $\widetilde{M}_y$. The solid (dashed) line marked by +(-) indicates the main (folded) band. **f**, The combination of nonsymmorphic and time reversal symmetries induces Weyl nodes on the invariant line. The red and blue solid (or dashed) lines indicate a pair of Kramer's bands. The labels indicate the eigenvalues of $\widetilde{M}_y$. **g,** Together with mirror symmetry $M_z$, all bands are two-fold degenerate and the essential Dirac point with four-fold degeneracy (marked by the black circle) occurs at π/a.